\documentclass[11pt,twoside]{article}

\usepackage{asp2004}
\usepackage{epsf}
\usepackage{psfig}
\usepackage{lscape}

\begin{document}
\title{Infrared Number Counts of AGN in the GOODS Fields}   
\author{Ezequiel Treister, C. Megan Urry and the GOODS AGN Team}   
\affil{Depto de Astronom\'{\i}a, U. de Chile, Santiago de Chile and Astronomy and Physics Departments, Yale University, New Haven, CT, USA}    

\begin{abstract} 
Using the known numbers and evolution of unobscured AGN, and
assuming that there is a large number of corresponding
obscured AGN --- as expected if AGN unification is correct
--- we predict the infrared number counts of AGN. These
agree well with the observed GOODS Spitzer data at 3.6, 8
and 24 $\mu$m, provided that the dust content of the
obscuring torus is $\sim 2$x less than assumed. This
strongly supports the AGN unification picture, and implies a
large population of infrared bright obscured AGN at all
redshifts and luminosities.
\end{abstract}


\section*{Introduction}	

To find obscured AGN beyond the local Universe requires the
combination of hard X-ray and infrared searches where most
of their emission is found. This was a strong motivation for
the Great Observatories Origins Deep Survey (GOODS), which
consists of deep high-resolution imaging from space in X
rays, far-infrared and optical, augmented with ground-based
imaging and spectroscopy. Here we discuss the infrared
properties of the AGN populations detected in X-rays in the
GOODS Fields. We explain the infrared number counts with a
simple unified model that includes many obscured AGN, as
required by population syntheses of the X-ray background. We
predict that only half the total sources are detected in
X-rays since they are heavily obscured. This population of
AGN missed in X-rays will be clearly visible in the Spitzer
data.

\section*{Model and Results}

Details of the AGN unification model used to predict the
observed number counts of AGN at different wavelengths were
presented by \citet{treister04}. Briefly, there are two main
ingredients for this model: (1) the AGN luminosity function
and its evolution \citep{ueda03}, which is based on hard
X-ray observations, and (2) the observed spectral energy
distributions (SEDs) of AGN, which vary with intrinsic
luminosity and neutral hydrogen column density ($N_H$) along
the line of sight. For the infrared part of the spectrum we
used the dust emission models of \citet{nenkova02}, varying
the geometry parameters within plausible limits, such that
local AGN SEDs are well fit and the ratio of obscured
($N_H>10^{22}$~cm$^{-2}$) to unobscured AGN is $\sim 3:1$ as
observed locally.

\begin{figure}
\plottwo{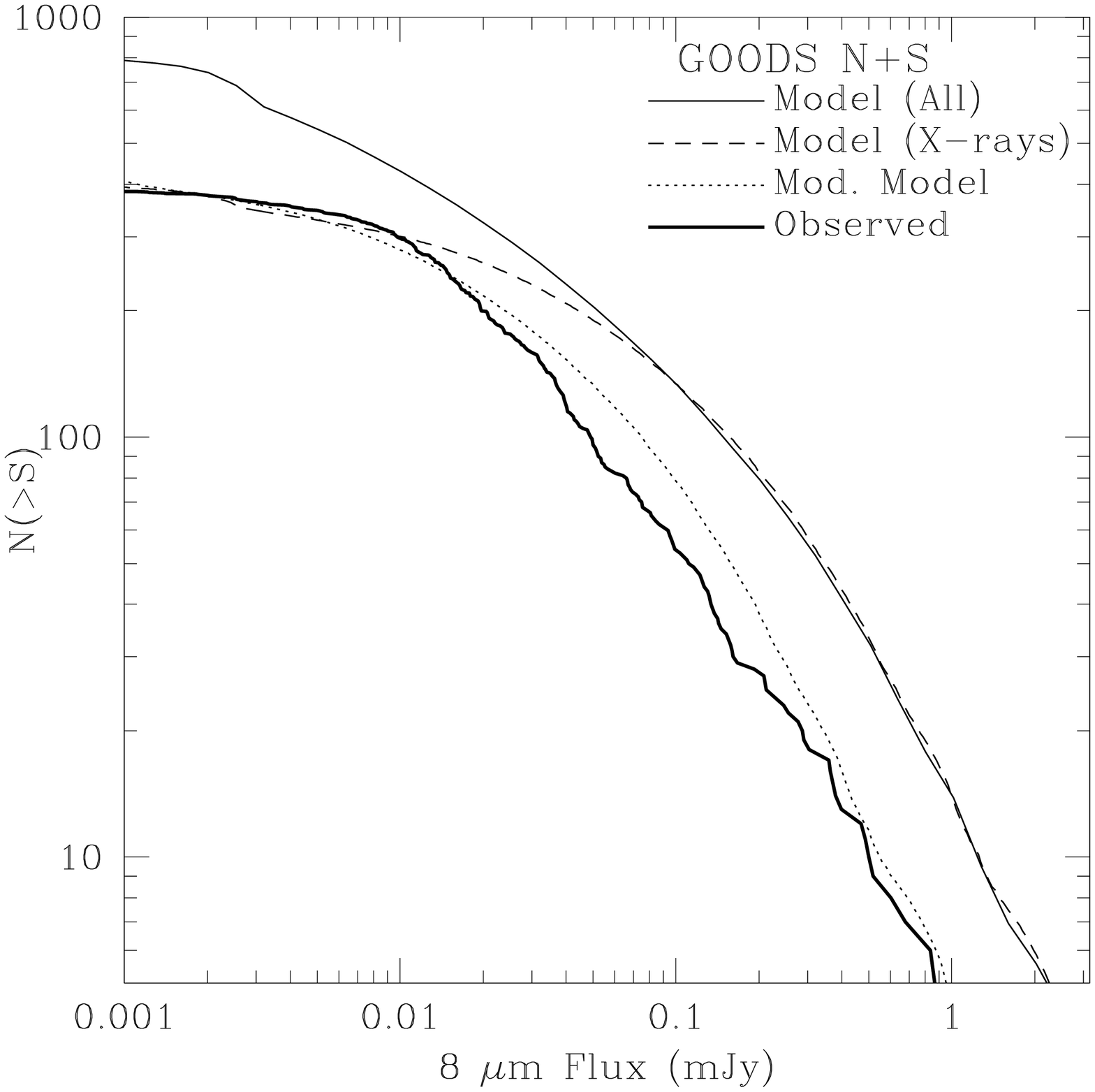}{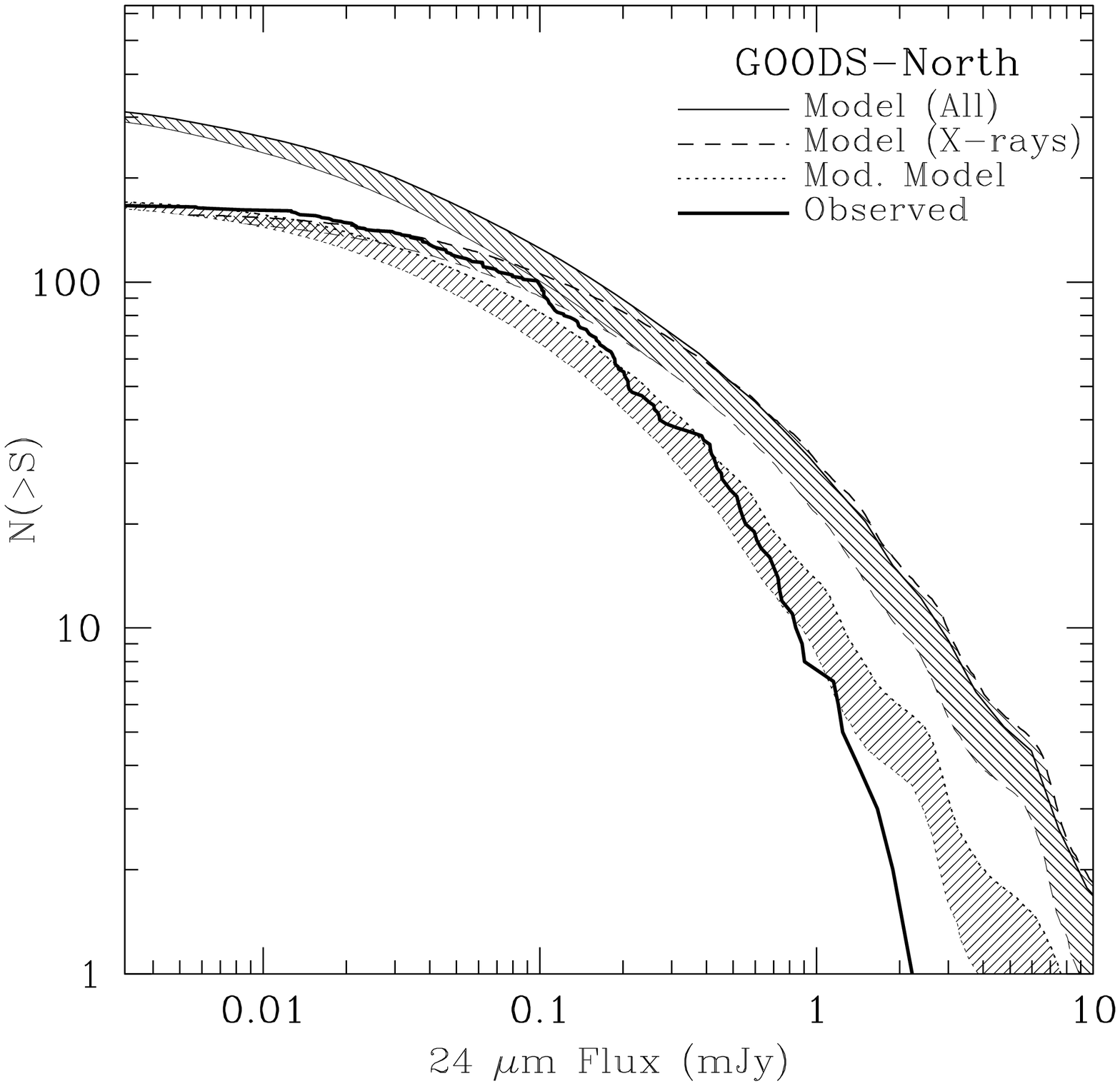}
\caption{Total ({\it thin solid line}) predicted AGN number 
counts for the GOODS-North and -South fields in the 8$\mu$m
band (left side) and for the North field only in the
24$\mu$m band (right side). Shaded regions in the 24$\mu$m
plot reflect the range of parameters for the IR spectra. The
{\it dashed line/region} show the prediction for AGN
detected in X-rays by Chandra. If the predicted fluxes are
reduced by $\sim 2$x ({\it dotted line/region}), the
predictions agree very well with the GOODS Spitzer
observations ({\it thick solid lines}).}
\end{figure}

In the 3.6 $\mu$m band, the observed distribution agrees
very well with the prediction of this model, i.e., with AGN
unification. These are the shortest wavelength observations
with Spitzer and thus trace the beginning of the torus
emission at z$\sim$0 and the host galaxy/AGN continuum at
higher redshifts. Therefore, they do not provide the
strongest test of AGN unification.

The predicted and observed number counts in the 8 and 24
$\mu$m Spitzer bands, shown in Fig.~1, are close to the
predictions but are a factor of $\sim 2$ lower in IR flux
(dotted lines in Fig.~1). Given that the number count
predictions depend on theoretical (and hence uncertain) dust
emission models, checked only against bright sources
observed by IRAS and ISO, we infer there may be slightly
less dust in the torus than previously assumed.

The model also predicts that only $\sim 50\%$ of the total
AGN in the GOODS fields are detected in deep Chandra X-rays
observations. The missed sources are heavily obscured AGN
and thus very faint in X-rays. All these AGN should be
detected in the deep Spitzer observations of these fields;
however the main challenge is to distinguish them from other
IR-bright sources like starburst galaxies.

\acknowledgements 

This work was supported in part by NASA grant
HST-GO-09425.13-A, Fundaci\'on Andes and the Centro de
Astrof\'{\i}sica FONDAP.



\begin{thebibliography}{}

\bibitem[{{Nenkova} {et~al.}(2002){Nenkova}, {Ivezi{\' c}}, \&
  {Elitzur}}]{nenkova02}
{Nenkova}, M., {Ivezi{\' c}}, {\v Z}., \& {Elitzur}, M. 2002, \apjl, 570, L9

\bibitem[Treister et al.(2004)]{treister04} Treister, E., et al.\ 
2004, \apj, 616, 123 

\bibitem[{{Ueda} {et~al.}(2003){Ueda}, {Akiyama}, {Ohta}, \& {Miyaji}}]{ueda03}
{Ueda}, Y., {Akiyama}, M., {Ohta}, K., \& {Miyaji}, T. 2003, \apj, 598, 886

\end{thebibliography}
\end{document}